\documentclass[aps,prl,twocolumn,superscriptaddress,showpacs,floatfix]{revtex4}
\usepackage{graphicx}
\usepackage{dcolumn}
\usepackage{bm}
\usepackage{amsfonts}
\usepackage{amsmath}
\def \ab {a_{\rm B}}

\begin{document}
\renewcommand{\thefigure}{\arabic{figure}}

\title{Theory of the Pseudospin Resonance in Semiconductor Bilayers}
\date{\today}

\author{Saeed H. Abedinpour}
\affiliation{NEST-CNR-INFM and Scuola Normale Superiore, I-56126 Pisa, Italy}
\author{Marco Polini}
\email{m.polini@sns.it}
\affiliation{NEST-CNR-INFM and Scuola Normale Superiore, I-56126 Pisa, Italy}
\author{A.H. MacDonald}
\affiliation{Department of Physics, The University of Texas at Austin, Austin, Texas 78712, USA}
\author{B. Tanatar}
\affiliation{Department of Physics, Bilkent University, Bilkent, 06800 Ankara, Turkey}
\author{M.P. Tosi}
\affiliation{NEST-CNR-INFM and Scuola Normale Superiore, I-56126 Pisa, Italy}
\author{G. Vignale}
\affiliation{Department of Physics and Astronomy, University of Missouri-Columbia, 
Columbia, Missouri 65211, USA}

\begin{abstract}
The pseudospin degree of freedom in a semiconductor bilayer gives rise to a collective mode analogous 
to the ferromagnetic resonance mode of a ferromagnet.  We present a theory of the dependence of the energy and the damping of 
this mode on layer separation $d$.  Based on these results, we discuss the possibility of realizing 
transport-current driven pseudospin-transfer oscillators in semiconductors.
\end{abstract}

\pacs{73.21.-b, 71.10.Ca, 76.50.+g, 85.75.-d}
\maketitle

\noindent {\it Introduction}---The layer degree of freedom in semiconductor bilayers is often regarded~\cite{ref:allan_1990} as 
an effective spin-$1/2$ pseudospin degree-of-freedom in which electrons in the top layer are assigned one pseudospin state,  
and electrons in the other layer the opposite one.  In the quantum Hall regime,
electron bilayers are sometimes~\cite{ref:allan_2004} pseudospin ferromagnets. 
The appearance of these broken-symmetry states motivates an interest in  
phenomena which are pseudospin analogs of the very robust magneto-electric effects 
which underpin spintronics in ferromagnetic metals, and therefore might underpin a useful pseudospintronic 
technology.  Unfortunately it appears likely~\cite{ref:gaetano} that pseudospin-ferromagnetism 
in semiconductor bilayers is a phenomenon that is limited to strong magnetic fields or, possibly, systems with 
extremely low densities.  This Letter is motivated by the observation that one important spintronic 
device, the spin-transfer oscillator~\cite{ref:silva_2004}, requires only collective spin-dynamics and not 
spontaneous magnetic order.  In a spin-transfer oscillator transport, currents drive
ferromagnetic-resonance collective spin-dynamics in 
the presence of applied fields strong enough to oppose hysteretic switching.  Since pseudospin paramagnets do support
a pseudospin resonance collective mode, as we explain below, pseudospin polarized transport currents in a semiconductor bilayer 
(easily realizable using individual-layer contacting techniques~\cite{ref:eisenstein_separate}) could,
provided that the resonance is sufficiently sharp, drive 
collective pseudospin dynamics and yield a device with similar functionality.
In this Letter we report on a theory of the damping of the pseudospin resonance which suggests that 
it is possible to design bilayers with sharp pseudospin resonances. 

In the absence of tunneling a semiconductor bilayer supports two types of collective excitations~\cite{ref:dassarma_madhukar,ref:experimental_references}: (i) an optical mode (the ordinary plasmon) with a long-wavelength dispersion relation $\propto q^{1/2}$ and weak damping  ($\propto q^2$),
in which electrons in the two layers oscillate in phase, and (ii) an acoustic plasmon with linear dispersion and strong Landau damping ($\propto q$)
by particle-hole excitations, in which the electrons in the two layers oscillate out of phase.  The pseudospin
resonance is a ${\bf q}=0$ collective mode
which develops from the out-of-phase plasmon when interwell tunneling is enabled.
In the pseudospin language interwell tunneling favors symmetric bilayer states and therefore acts like a pseudospin-magnetic field,
which we take to act in the $\hat z$ direction.  The pseudospin resonance then involves collective precession around this pseudospin field, with $\hat y$-direction pseudospins representing current flowing between the layers and $\hat x$-direction 
pseudospins representing charge accumulation in one of the layers. Theoretical treatments of the pseudospin resonance 
have so far relied on the random phase approximation~\cite{ref:dassarma} (RPA), sometimes with Hubbard or local-density-approximation
corrections~\cite{ref:dassarma,ref:bolcatto}.  These papers demonstrate that interactions shift the resonance away from the position of
the ${\bf q}=0$ particle-hole excitations, eliminating the Landau damping process. 
In this Letter we present a theory of the pseudospin transfer resonance that is based on a systematic expansion in powers
of the difference $V_{-}$ between intra- and inter-layer electron-electron interaction.  We obtain an expression for the interaction-induced 
resonance position shift which is exact to leading order in $V_{-}$, and an expression for the leading order damping contributions which 
appears at second order in $V_{-}$.  Damping of the pseudospin resonance is similar to damping of the ferromagnetic
resonance~\cite{ref:gilbert,ref:evelina_2006} in a metal, except that it is intrinsic and driven by 
electron-electron interactions rather than disorder. The physical mechanism of damping is the production of two (or more) electron-hole pairs 
with zero total momentum. The phase space for these processes implies that the damping rate is proportional to the 
cube of the resonance frequency, implying that the resonance is sharp whenever its energy is small compared to the 
Fermi energy of the bilayer system. 

\noindent {\it The model}---  In a semiconductor bilayer, electrons in the same layer
interact through the two-dimensional ($2D$) Coulomb interaction $V_{\rm s}(q)=2\pi e^2/(\epsilon q)$ ($\epsilon$ is the dielectric constant), while electrons in different layers are coupled through the interlayer Coulomb interaction $V_{\rm d}(q)=V_{\rm s}(q)e^{-qd}$. 
We assume a spatially constant inter-layer tunneling amplitude which we denote by  
$\Delta_{\rm SAS}/2$ and present our theory using a pseudospin representation in which the tunneling term is diagonal,
{\em i.e.} the representation in which $|\!\!\uparrow\rangle$  refers to the symmetric combination of 
single-layer states and $|\!\!\downarrow\rangle$ to the antisymmetric combination.  
The total Hamiltonian is then ($\hbar=1$)
\begin{eqnarray}\label{eq:spin_spin}
{\hat {\cal H}}&=&
-\Delta_{\rm SAS}{\hat S}^z_{\rm tot}+\sum_{{\bf k}, \alpha, \sigma}
\frac{{\bf k}^2}{2m}{\hat c}^{\dagger}_{{\bf k}, \alpha, \sigma}{\hat c}_{{\bf k}, \alpha, \sigma}
\nonumber\\
&+&\frac{1}{2S}
\sum_{{\bf q}\neq 0}V_+(q){\hat \rho}_{\bf q}{\hat \rho}_{-{\bf q}}+
\frac{2}{S} \sum_{{\bf q}\neq 0}V_-(q){\hat S}^x_{\bf q}{\hat S}^x_{-{\bf q}}\,.
\end{eqnarray}
Here $\sigma$ is the real-spin label, $\alpha$ is the pseudospin label,  $S$ is the sample area, 
${\hat \rho}_{\bf q}=\sum_{{\bf k}, \alpha, \sigma}{\hat c}^{\dagger}_{{\bf k}-{\bf q}/2, \alpha, \sigma}{\hat c}_{{\bf k}+{\bf q}/2, \alpha, \sigma}$ 
and ${\hat S}^a_{\bf q}=\sum_{{\bf k}, \alpha, \beta, \sigma} 
{\hat c}^{\dagger}_{{\bf k}-{\bf q}/2, \alpha, \sigma}(\tau^a_{\alpha\beta}/2){\hat c}_{{\bf k}+{\bf q}/2, \beta, \sigma}$ are the total density and the pseudospin operators 
($\tau^a$ being Pauli matrices with $a=x,y,z$), 
${\hat S}^a_{\rm tot}= {\hat S}^a_{{\bf q}=0}$, and, finally, $V_\pm(q)=[V_{\rm s}(q)\pm V_{\rm d}(q)]/2$. 

\noindent {\it Theory}---The theory we develop in this Letter is based on the observation that the difference between the intra- and inter-layer interaction $V_-(q) =\pi e^2 (1 -e^{-qd})/(\epsilon q)$ is always smaller than  $\pi e^2 d/\epsilon$, which becomes a small perturbation when  $d \ll {\rm max}(r_s \ab, \ab/r_s^2)$. Here $r_s=(\pi n a^2_{\rm B})^{-1/2}$ is the Wigner-Seitz density parameter and $a_{\rm B}=\epsilon/(m e^2)$ is  the Bohr radius.  
The above inequality guarantees that the third term in the Hamiltonian (\ref{eq:spin_spin}) is a small perturbation either compared to  
the kinetic energy [$\sim e^2/(r^2_s \epsilon a_{\rm B})$] which dominates in the high-density limit, or compared to the interaction energy
[$\sim e^2/(r_s \epsilon a_{\rm B})$] which dominates in the low-density limit. We will therefore perform a systematic expansion for the pseudospin 
resonance frequency and damping rate in powers of $V_{-}(q)$.  Our approach will be asymptotically exact in the limit $d \ll \ab$,
and is expected to be qualitatively correct for $d \sim \ab$.   

We determine the properties of the pseudospin resonance by evaluating the 
transverse pseudospin response function $\chi_{S^x S^x}(q, \omega)=\langle\langle {\hat S}^x_{{\bf q}};{\hat S}^x_{-{\bf q}}\rangle\rangle_\omega/S$, 
where we have introduced the Kubo product 
$\langle\langle {\hat A}; {\hat B}\rangle\rangle_\omega
=-i\lim_{\epsilon\rightarrow 0^+}\int_0^{+\infty}dt\, e^{i\omega t}e^{-\epsilon t}
\langle \Psi_{\rm GS}|[{\hat A}(t), {\hat B}(0)]|\Psi_{\rm GS}\rangle$~\cite{ref:giuliani_and_vignale}. Since our aim is to calculate the transverse mode at $q=0$ we will focus on the response function $\chi_{S^xS^x}(\omega)\equiv \chi_{S^xS^x}(q=0,\omega)$~\cite{footnote1}.
The in-plane pseudospin operators satisfy the Heisenberg equations of motion, 
\begin{equation}\label{eq:spin_dynamics}
\left\{
\begin{array}{l}
{\displaystyle \partial_t {\hat S}^x_{\rm tot}=\Delta_{\rm SAS}{\hat S}^y_{\rm tot}}
\vspace{0.1 cm}\\
{\displaystyle \partial_t {\hat S}^y_{\rm tot}=-\Delta_{\rm SAS}{\hat S}^x_{\rm tot}-\frac{2}{S}\sum_{{\bf k}} V_-(k)({\hat S}^z_{\bf k}{\hat S}^x_{-{\bf k}}+{\hat S}^x_{{\bf k}}{\hat S}^z_{-{\bf k}})}
\end{array}
\right.\,;
\end{equation}
${\hat S}^x_{\rm tot}$, which measures the difference between charges in the two layers, is a good quantum number 
when $\Delta_{\rm SAS} \to 0$ whereas ${\hat S}^y_{\rm tot}$ is not conserved even in this limit because of 
the pseudospin-dependent interactions.  When $d \to 0$ these equations reduce to a pseudospin version
of Larmor's theorem, in which the precession is undamped and its frequency is given exactly by the non-interacting particle value $\Delta_{\rm SAS}$. 

Our theory starts by making repeated use of Eqs.~(\ref{eq:spin_dynamics}) in  
the Kubo product identity~\cite{ref:giuliani_and_vignale,ref:mdt}: 
$
\langle\langle {\hat A};{\hat B}\rangle\rangle_\omega=\langle \Psi_{\rm GS}|[{\hat A},{\hat B}] |\Psi_{\rm GS}\rangle/\omega+i
\langle\langle \partial_t {\hat A};{\hat B}\rangle\rangle_\omega/\omega\,.
$ 
After some algebraic manipulations 
we arrive at the following {\it exact} expression for $\chi_{S^xS^x}(\omega)$
\begin{widetext}
\begin{eqnarray}\label{eq:chi_xx_final}
\chi_{S^xS^x}(\omega)
={\cal M}^z\frac{\Delta_{\rm SAS}}{\Omega^2}+
\frac{4\Delta_{\rm SAS}^2}{\Omega^4 S^2}
\sum_{{\bf k}} V_-(k)f({\bf k})+\frac{2i\omega\Delta_{\rm SAS}}{\Omega^4 S^2}\sum_{{\bf k}} V_-(k)g({\bf k})
+\frac{4\Delta^2_{\rm SAS}}{\Omega^4 S^3}\sum_{{\bf k},{\bf k}'} V_-(k)V_-(k'){\cal L}({\bf k}, {\bf k}', \omega)\,.
\end{eqnarray}
\end{widetext}
Here $\Omega^2=\omega^2-\Delta^2_{\rm SAS}$, ${\cal M}^z=\langle \Psi_{\rm GS}|{\hat S}^z_{\rm tot}|\Psi_{\rm GS}\rangle/S$ is the ground-state pseudospin magnetization per unit area, $f({\bf k})=\langle \Psi_{\rm GS}|{\hat S}^z_{\bf k}{\hat S}^z_{-{\bf k}}|\Psi_{\rm GS}\rangle-
\langle \Psi_{\rm GS}|{\hat S}^x_{\bf k}{\hat S}^x_{-{\bf k}}|\Psi_{\rm GS}\rangle$, $g({\bf k})=\langle \Psi_{\rm GS}|{\hat S}^x_{\bf k}{\hat S}^y_{-{\bf k}}|\Psi_{\rm GS}\rangle+
\langle \Psi_{\rm GS}|{\hat S}^y_{\bf k}{\hat S}^x_{-{\bf k}} |\Psi_{\rm GS}\rangle$, and ${\cal L}({\bf k}, {\bf k}', \omega)=
\langle\langle [{\hat S}^z_{\bf k}{\hat S}^x_{-{\bf k}}+{\hat S}^x_{\bf k}{\hat S}^z_{-{\bf k}}]; [{\hat S}^z_{{\bf k}'}{\hat S}^x_{-{\bf k}'}+{\hat S}^x_{{\bf k}'}{\hat S}^z_{-{\bf k}'}]\rangle\rangle_\omega$. Notice that $f({\bf k})$ is purely real, $g({\bf k})$ is purely imaginary, and ${\cal L}({\bf k}, {\bf k}', \omega)$ has both a real and an imaginary part. 
The symmetric interaction $V_+$ does not appear explicitly in Eq.~(\ref{eq:chi_xx_final}). 
When $V_{-}$ is set to zero ($d \to 0$), the interaction part of the Hamiltonian is pseudospin invariant.  
Larmor's theorem then applies to the pseudospin degree-of-freedom and only the first term on the right 
hand-side of Eq.~(\ref{eq:chi_xx_final}) survives.  
We refer to the Hamiltonian ${\hat {\cal H}}$ at $V_-=0$ as the {\it reference system} (RS),
on which the perturbative scheme outlined below is based.  

The key idea now is to expand $\chi_{S^xS^x}(\omega)$ in powers of $V_-$. 
For example, the ground-state pseudospin magnetization ${\cal M}^z$ is expanded as ${\cal M}^z={\cal M}^z_0+{\cal M}^z_1+{\cal M}^z_2+...$, where the $n$-th term ${\cal M}^z_n$ is ${\cal O}(V^n_{-})$. The quantities $f$, $g$, and ${\cal L}$ are similarly expanded. 
Note that the zero-th order of $f({\bf k})$, denoted by $f_0({\bf k})$,  is a non-zero difference between longitudinal and 
transverse pseudospin structure factors.
On the other hand, the zero-th order of  $g({\bf k})$, denoted by $g_0({\bf k})$ 
vanishes because the RS Hamiltonian is invariant under rotations by $90$ degrees about the $\hat{z}$-axis in pseudospin space
which map ${\hat S}_x \to {\hat S}_y$ and ${\hat S}_y \to - {\hat S}_x$.   

The pseudospin resonance frequency is the solution of the equation $ \Re e [\chi^{-1}_{S^xS^x}(\omega_\perp)]=0$. 
The inverse of  $\chi_{S^xS^x}$ can be expanded with the help of the formula $\chi^{-1}_{S^xS^x}(\omega)=\chi^{-1}_0(\omega)-\chi^{-2}_0(\omega)\chi_1(\omega)+\chi^{-2}_0(\omega)[\chi^{-1}_0(\omega)\chi^2_1(\omega)-\chi_2(\omega)]+...$, where $\chi_n(\omega)$ is the $n$-th order term in the expansion for  $\chi_{S^xS^x}$. To appreciate the power of Eq.~(\ref{eq:chi_xx_final}) we first use it to find $\omega_\perp$ to first order in ${\bar d}=d/a_{\rm B}$.  Keeping only terms up to first order in Eq.~(\ref{eq:chi_xx_final}) and making use of the formula above for the power series of $\chi^{-1}_{S^xS^x}(\omega)$ we obtain
\begin{eqnarray}\label{eq:first_order}
\chi^{-1}_{S^xS^x}(\omega)&=&\frac{\omega^2-\Delta_{\rm SAS}^2}{{\cal M}^z_0\Delta_{\rm SAS}}\left(1-\frac{{\cal M}^z_1}{{\cal M}^z_0}\right)\nonumber\\
&-&\frac{4V_-(0)}{({\cal M}^z_0)^2}
\frac{1}{S^2}\sum_{{\bf k}} f_0({\bf k})+{\cal O}({\bar d}^2)\,,
\end{eqnarray}
which implies immediately that 
 \begin{equation}\label{eq:plasmon_1st_order}
\omega^2_\perp=\Delta^2_{\rm SAS}+\frac{4 \Delta_{\rm SAS} V_-(0)}{{\cal M}^z_0}\frac{1}{S^2}\sum_{{\bf k}}f_0({\bf k})+{\cal O}({\bar d}^2)\,.
\end{equation}
This equation is exact to all orders in the intralayer Coulomb interaction $V_{\rm s}$.
In the high-density (non-interacting) limit one can find simple analytical expressions for ${\cal M}^z_0$ and $f_0({\bf k})$, ${\cal M}^z_0=(n_{\rm S}-n_{\rm AS})/2$ and $S^{-2}\sum_{{\bf k}}f_0({\bf k})=({\cal M}^z_0)^2/2$. Here $n_\alpha=k^2_{\rm F\alpha}/(2\pi)$ are the band occupation factors, $k_{\rm F\alpha}$ being 
the Fermi wavenumber for band $\alpha$. 
In this limit Eq.~(\ref{eq:plasmon_1st_order}) simplifies to $\omega^2_\perp=\Delta^2_{\rm SAS}+2\Delta_{\rm SAS}{\cal M}^z_0V_-(0)+{\cal O}({\bar d}^2)$. The second term, which supplies the interaction induced shift in the 
pseudospin resonance position, is a factor of two smaller than in RPA theory~\cite{ref:dassarma}. 
The source of this difference is  easy to understand: our calculation includes the first-order exchange corrections 
to the resonance frequency which are absent in the RPA.  Since $V_{-}$ is independent of $q$ at first order in $d$,
corresponding to a $\delta$-function interaction in real space, the like-real-spin contribution to the resonance position shift 
present in the RPA is canceled by exchange interactions.    

\begin{figure}
\begin{center}
\includegraphics[width=0.8\linewidth]{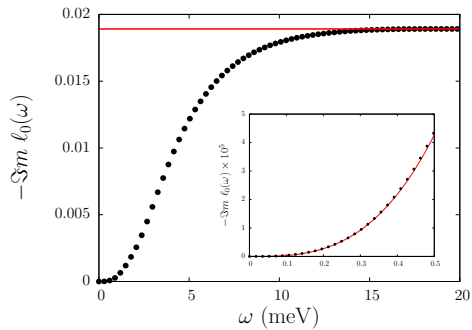}
\caption{(Color online) Imaginary part of the dynamical response function $\ell_0(\omega)$ 
(in units of ${\rm eV}^{-1}{\rm nm}^{-6}$) as a function of $\omega$ for a bilayer electron gas with $n=8.3\times 10^{10}~{\rm cm}^{-2}$ and 
$\Delta_{\rm SAS}=1.48~{\rm meV}$. The $\delta$-function contribution at $\omega=\Delta_{\rm SAS}$ [see first term in Eq.~(\ref{eq:4point_response})] has been subtracted. The red solid line is the asymptotic result $\Im m~\ell_0(\omega\to \infty)=-m n^2/32$. 
Inset: a zoom of the low-energy region. The solid red curve is the expression 
$\Im m~\ell_0(\omega)=-\gamma \omega^3$ with $\gamma \simeq 3.41\times 10^{-4}$~\cite{ref:gamma}.}
\label{fig:one}
\end{center}
\end{figure}

The main object of this work is to estimate the resonance decay rate, which appears first at second-order 
in $V_{-}$ and is zero in the RPA.  (Additional interaction corrections to the resonance position $\Re e(\omega_\perp)$, 
which we do not discuss at length, also appear at second order~\cite{footnote2}.)  
The linewidth of the pseudospin resonance [$=-2\Im m (\omega_\perp)$] is given, up to second order in $d$, by
\begin{equation}\label{eq:gamma}
\Gamma_\perp=-\frac{4V^2_-(0)\Delta_{\rm SAS}}{{\cal M}^z_0}\times
\lim_{\omega\rightarrow \Delta_{\rm SAS}}\frac{\Im m~\ell_0(\omega)}{\omega}\,.
\end{equation}
where $\ell_0(\omega)$ is the wavevector sum of the four-spin correlation function ${\cal L}({\bf k}, {\bf k}', \omega)$.
This quantity can be evaluated analytically in the high-density kinetic-energy dominated limit
in which it is dominated by a decay process in which two particle-hole pairs 
are excited out of the Fermi sea, one involving a pseudospin-flip.  The second
particle-hole excitation is diagonal in pseudospin and absorbs the momentum emitted by the first.
We find that
\begin{widetext}
\begin{eqnarray}\label{eq:4point_response}
\Im m~\ell_0(\omega)=-\frac{\pi}{2}({\cal M}^z_0)^3 \delta(\omega-\Delta_{\rm SAS})-\frac{\pi}{2 S^3}\!\!\sum_{{\bf k}, {\bf k}', {\bf k}''}
\sum_{\alpha, \beta}\delta(\omega-\Omega_{\alpha}({\bf k},{\bf k}'))n_{{\bf k}'',\alpha}n_{{\bf k}''-{\bf k}+{\bf k}',\beta}
(1-n_{{\bf k}''+{\bf k}',\beta})(1-n_{{\bf k}''-{\bf k},{\bar \alpha}})
\end{eqnarray}
\end{widetext}
where $n_{{\bf k},\alpha}=\Theta(k_{\rm F\alpha}-|{\bf k}|)$ and $\Omega_{\alpha}({\bf k},{\bf k}')={\bf k}\cdot {\bf k}'/m+\alpha \Delta_{\rm SAS}$. 
The first term on the right-hand-side of Eq.~(\ref{eq:4point_response}) does not contribute to the life-time of the shifted resonance.
In Fig.~\ref{fig:one} we illustrate the dependence of $\Im m~\ell_0(\omega)$ on $\omega$. 
The $\omega^3$ dependence at small $\omega$ is the double-particle-hole excitation 
manifestation of the familiar Pauli-blocking reduction in the excitation density-of-states 
in a Fermi sea which underlies Fermi liquid theory; damping drops much more rapidly
at low energies than for ferromagnetic resonance~\cite{ref:evelina_2006}
dominated by single-particle decay processes.
Eqs.~(\ref{eq:chi_xx_final}), (\ref{eq:plasmon_1st_order}), (\ref{eq:gamma}), and (\ref{eq:4point_response}) constitute the most important results of this work and provide, to best of our knowledge, the first microscopic theory of the pseudospin resonance linewidth. 

\begin{figure}
\begin{center}
\includegraphics[width=0.8\linewidth]{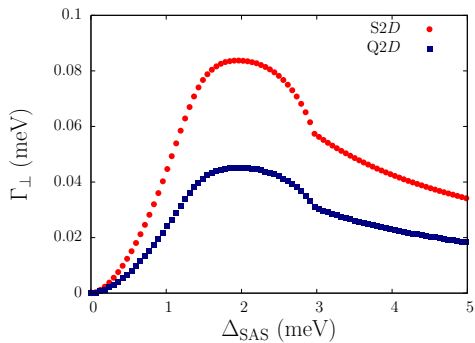}
\caption{(Color online) Intrinsic linewidth $\Gamma_\perp$ of the pseudospin resonance as a function of $\Delta_{\rm SAS}$ for a bilayer with density $n=8.3\times 10^{10}~{\rm cm}^{-2}$ and $d=50$ \AA. 
The S$2D$ curve was evaluated using the bare $2D$ interactions $V_{\rm s}(q)$ and $V_{\rm d}(q)$ defined above
whereas the Q$2D$ result was evaluated with more realistic interactions weakened by form factors~\cite{form_factors} which account for typical
quantum well widths.
}
\label{fig:two}
\end{center}
\end{figure}

\begin{figure}
\begin{center}
\includegraphics[width=0.8\linewidth]{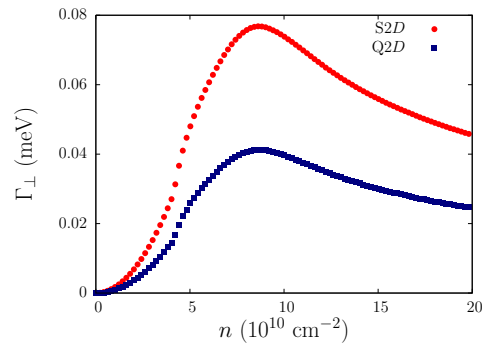}
\caption{(Color online) Intrinsic linewidth $\Gamma_\perp$ of the pseudospin resonance as a function of $n$ 
for a bilayer with tunneling gap $\Delta_{\rm SAS}=1.48~{\rm meV}$ and $d=50$ \AA. The labels S$2D$ and Q$2D$ have the same meaning as in Fig.~\ref{fig:two}.}
\label{fig:three}
\end{center}
\end{figure}

\noindent {\it Numerical results and discussion}--- Typical numerical results for $\Gamma_\perp$, 
calculated from Eqs.~(\ref{eq:gamma}) and~(\ref{eq:4point_response}) are shown in Figs.~\ref{fig:two} and~\ref{fig:three}.
In Fig.~\ref{fig:two} we show $\Gamma_\perp$ as a function of $\Delta_{\rm SAS}$ for a bilayer 
with density $n=8.3\times 10^{10}~{\rm cm}^{-2}$ and interlayer distance 
$d=L+w=50$~\AA. Here $L=40$~\AA~is the width of each quantum well and $w=10$~\AA~
is the barrier width (we have chosen material parameters corresponding to a GaAs/AlGaAs bilayer).
The non-analytic behavior of $\Gamma_\perp$ for $\Delta_{\rm SAS}\sim 3~{\rm meV}$ is due to the transition
from the situation in which both symmetric and antisymmetric bands are occupied to that 
in which only the symmetric band is occupied. In Fig.~\ref{fig:three} we illustrate 
the dependence of $\Gamma_\perp$ on density for a fixed value of  $\Delta_{\rm SAS}=1.48~{\rm meV}$. 
Since the resonance frequency is close to $\Delta_{\rm SAS}$, these calculations predict that 
the pseudospin resonance can be very sharp, especially when $\Delta_{\rm SAS}$ is small compared to the 
Fermi energy of the bilayer.  On physical grounds we expect that the main effect of going to higher 
order in $d$ will be to replace the bare interlayer interaction in Eq.~(\ref{eq:gamma}) by a weaker screened interaction,
further reducing the damping. 

Our theory of the resonance amounts to the derivation of an {\it anisotropic}, linearized pseudospin Landau-Liftshitz-Slonczewski equation: 
\begin{equation}\label{eq:LLS}
\left\{
\begin{array}{l}
{\displaystyle \partial_t {\cal M}^x=\Delta_{\rm SAS} {\cal M}^y-\frac{I}{e}}
\vspace{0.1 cm}\\
{\displaystyle \partial_t {\cal M}^y= -\frac{\omega_{\perp}^2+\Gamma^2_\perp}{\Delta_{\rm SAS}}{\cal M}^x- 
\frac{2\Gamma_{\perp}}{\Delta_{\rm SAS}} \partial_t {\cal M}^x}
\end{array}
\right.\,;
\end{equation}
where ${\cal M}^a$ is the average macroscopic pseudospin polarization, which becomes equal to $\langle {\hat S}^{a}_{\rm tot} \rangle$ in the limit $I \to 0$. In the first line of Eq.~(\ref{eq:LLS}) we have added a Slonczewski~\cite{ref:slonczewski} pseudospin transfer term
proportional to the tunnel current $I$, which is injected in one layer and extracted from the other.
As in the ferromagnetic case, it is the reaction counterpart of the torque which acts on the 
transport quasiparticles to enable their transfer between layers upon moving through the sample, 
and must be present because of the nearly exact conservation of pseudospin by interactions.
In the second line of Eq.~(\ref{eq:LLS}) we have added a Gilbert-like damping term $\propto \partial_t {\cal M}_x$ 
(the anisotropy of the Gilbert damping in the present problem derives from the strongly anisotropic character 
of the interaction part of the Hamiltonian).
These equations 
[which describe a damped pseudospin precession of frequency $\omega_\perp$ and damping rate $\Gamma_\perp$ 
about the steady state values 
${\cal M}^y(t\to \infty)=I /(e\Delta_{\rm SAS})$, 
${\cal M}^x(t\to \infty)=0$] 
are similar to those which describe
spin-transfer torque oscillators~\cite{ref:silva_2004} in ferromagnets and suggest that similar, and possibly
more flexible, devices could be realized in semiconductor bilayers.  We anticipate that the pseudospin resonance 
in ferromagnets will have negative rather than positive dispersion, because of the $q$ dependence of $V_{-}(q)$.
The roles of this property, and the fact that the single-particle and collective excitation frequencies are not 
widely separated, are difficult to fully anticipate.  Nevertheless, this work suggests that experimental 
studies of non-linear transport in bilayers have great potential.

\noindent {\it Acknowledgments}--- We thank Vittorio Pellegrini for helpful discussions. 
M.P. acknowledges the hospitality of the Department of Physics and Astronomy of the University of Missouri-Columbia. 
A.H.M. was supported by the Welch Foundation, the ARO, and SWAN-NRI.  
G.V. was supported by NSF Grant No. DMR-031368.


\begin{references}
\bibitem{ref:allan_1990}
	A.H. MacDonald, P.M. Platzman, and G.S. Boebinger, \prl {\bf 65}, 775 (1990).
\bibitem{ref:allan_2004}
	J.P. Eisenstein and A.H. MacDonald, Nature {\bf 432}, 691 (2004).
\bibitem{ref:gaetano}
	S. Conti and G. Senatore, Europhys. Lett. {\bf 36}, 695 (1996);
	L. Zheng, M.W. Ortalano, and S. Das Sarma, \prb {\bf 55}, 4506 (1997).
\bibitem{ref:silva_2004}
        S.I. Kiselev {\em et al.}, Nature {\bf 425}, 380 (2003);
	W.H. Rippard {\em et al.},  \prl {\bf 92}, 027201 (2004); 
	A.A. Tulapurkar {\em et al.}, Nature {\bf 438}, 339 (2005).
\bibitem{ref:eisenstein_separate} 
	J.P. Eisenstein, L.N. Pfeiffer, and K.W. West, \apl {\bf 57}, 2324 (1990).
\bibitem{ref:dassarma_madhukar}
	S. Das Sarma and A. Madhukar, \prb {\bf 23}, 805 (1981);
	G.E. Santoro and G.F. Giuliani, {\it ibid.} {\bf 37}, 937 (1988).	
\bibitem{ref:experimental_references}
	R. Decca {\it et al.}, \prl {\bf 72}, 1506 (1994);
	A.S. Plaut {\it et al.}, \prb {\bf 55}, 9282 (1997);
	D.S. Kainth {\it et al.}, J. Phys.: Condens. Matter {\bf 12}, 439 (2000) and \prb {\bf 59}, 2095 (1999);
	S. Holland {\it et al.}, \prb {\bf 66}, 073305 (2002).
\bibitem{ref:dassarma}
	S. Das Sarma and E.H. Hwang, \prl {\bf 81}, 4216 (1998). 
	Notice that within this RPA calculation the frequency of transverse plasmon at $q=0$ is a strictly linear function of the interlayer separation $d$.
\bibitem{ref:bolcatto}	
	P.G. Bolcatto and C.R. Proetto, \prl {\bf 85}, 1734 (2000). 
\bibitem{ref:gilbert}
	See for example 
	V. Korenman and R.E. Prange, \prb {\bf 6}, 2769 (1972);
	J. Sinova {\it et al.}, \prb {\bf 69}, 085209 (2004);
	Y. Tserkovnyak, A. Brataas, G.E.W. Bauer, and B.I. Halperin, \rmp {\bf 77}, 1375 (2005) and other related works cited in these papers.
\bibitem{ref:evelina_2006}	
	E.M. Hankiewicz, G. Vignale, and Y. Tserkovnyak, \prb {\bf 75}, 174434 (2007).
\bibitem{ref:giuliani_and_vignale}
	G.F. Giuliani and G. Vignale, {\it Quantum Theory of the Electron Liquid} 
	(Cambridge University Press, Cambridge, 2005).
\bibitem{footnote1}
	The other in-plane pseudospin susceptibilities are completely determined by
	$\chi_{S^xS^x}(\omega)$: we have
	$
	\chi_{S^yS^x}(\omega)=-i\omega\chi_{S^xS^x}(\omega)/\Delta_{\rm SAS}
	$
	and
	$
	\chi_{S^yS^y}(\omega)=-{\cal M}/\Delta_{\rm SAS}+
	\omega^2\chi_{S^xS^x}(\omega)/\Delta^2_{\rm SAS}
	$, where ${\cal M}$ is defined after Eq.~(\ref{eq:chi_xx_final}).
\bibitem{ref:mdt}	
	R. Nifos\`i, S. Conti, and M.P. Tosi, \prb {\bf 58}, 12758 (1998); 
	I. D'Amico and G. Vignale, {\it ibid.} {\bf 62}, 4853 (2000); Z. Qian and G. Vignale, \prl {\bf 88}, 056404 (2002).
\bibitem{footnote2}
	The expression for the frequency of the transverse pole up to order ${\bar d}^2$ is cumbersome and will be presented elsewhere.
\bibitem{ref:gamma}
	The analytical expression for $\gamma$ is rather cumbersome and will be presented elsewhere.	
\bibitem{form_factors}
	See {\it e.g.} Eqs.~(15) and~(16) in R. C\^ot\'e, L. Brey, and A.H. MacDonald, \prb {\bf 46}, 10239 (1992).
\bibitem{ref:slonczewski} 
	J.A. Slonczewski, J. Magn. Magn. Mater. {\bf 159}, L1 (1996).
\end{references}
\end{document}